\documentclass[aps,prb,twocolumn,groupedaddress,showpacs,floatfix,
               superscriptaddress, nofootinbib]{revtex4-2}

\usepackage[colorlinks = true,
            linkcolor = red,
            urlcolor  = red,
            citecolor = blue,
            anchorcolor = blue]{hyperref}
\usepackage{epsfig}
\usepackage{amsmath,amssymb}
\usepackage{physics}
\usepackage{graphicx}
\usepackage[dvipsnames,usenames]{xcolor}

\usepackage[normalem]{ulem}
\usepackage{todonotes}
\usepackage{soul} 
\usepackage{orcidlink}
\usepackage{glossaries}
\newacronym{TRIQS}{TRIQS}{Toolbox for Research on Interacting Quantum Systems}
\newacronym{DMFT}{DMFT}{dynamical mean field theory}
\newacronym{ML}{ML}{machine learning}
\newacronym{QMC}{QMC}{quantum Monte Carlo}
\newacronym{QMCacc}{QMC-Acc}{neural network accelerated quantum Monte Carlo}
\newacronym{MAE}{MAE}{mean absolute error}
\newacronym{SLMC}{SLMC}{self-learning Monte Carlo}
\newacronym{AIM}{AIM}{Anderson impurity model}
\newacronym{ED}{ED}{exact diagonalization}
\newacronym{CTQMC}{CT-QMC}{continuous-time QMC}
\newacronym{CTQMC_Long}{CT-QMC}{continuous-time quantum Monte Carlo}
\newacronym{NRG}{NRG}{numerical renormalization group}
\newacronym{NN}{NN}{neural network}
\newacronym{CTHYB}{CTHYB}{continuous-time hybridization-expansion}
\newacronym{RMSE}{RMSE}{root-mean-square error}
\newacronym{NNQIS}{NN-QIS}{neural network quantum impurity solver}

\newcommand{\UcI}{U_{\mathrm{c}1}}
\newcommand{\UcII}{U_{\mathrm{c}2}}

\begin{document}
\preprint{}
\title{Neural networks as low-cost surrogates for impurity solvers in quantum embedding methods}

\author{Rohan Nain\orcidlink{0000-0002-8476-9567}}
\affiliation{Department of Physics and Astronomy, The University of Tennessee, Knoxville, Tennessee 37996, USA}
\affiliation{Computational Sciences and Engineering Division, Oak Ridge National Laboratory, Oak Ridge, Tennessee, 37831-6494, USA}

\author{Philip M. Dee\orcidlink{0000-0002-4249-9036}}
\affiliation{Computational Sciences and Engineering Division, Oak Ridge National Laboratory, Oak Ridge, Tennessee, 37831-6494, USA}

\author{Kipton Barros\orcidlink{0000-0002-1333-5972}}
\affiliation{Theoretical Division and CNLS, Los Alamos National Laboratory, Los Alamos, New Mexico 87545, USA} 

\author{Steven~Johnston\orcidlink{0000-0002-2343-0113}}
\affiliation{Department of Physics and Astronomy, The University of Tennessee, Knoxville, Tennessee 37996, USA}
\affiliation{Institute for Advanced Materials and Manufacturing, The University of Tennessee, Knoxville, Tennessee 37996, USA} 

\author{Thomas A. Maier\orcidlink{0000-0002-1424-9996}}
\affiliation{Computational Sciences and Engineering Division, Oak Ridge National Laboratory, Oak Ridge, Tennessee, 37831-6494, USA}

\begin{abstract}

A promising application of machine learning is the creation of low-cost surrogate models to mitigate computational bottlenecks in quantum many-body simulations. Here, we explore whether a \gls*{NN} can be trained in the low-data regime, with one to two orders of magnitude fewer training examples than previous works, as an efficient substitute for the impurity solver in dynamical mean-field theory simulations of correlated electron models. We show that the \gls*{NN} solver achieves accuracy comparable to popular \gls*{CTQMC_Long} impurity solvers when interpolating between samples within the training set. While the \gls*{NN}'s performance decreases notably when extrapolating to lower temperatures outside the training distribution, its output still provides an excellent initial guess for input to more accurate \gls*{CTQMC_Long} impurity solvers, thus accelerating the time to solution up to a factor of five.  We discuss our results in the context of rapid phase-space exploration. 
\end{abstract} 

\glsresetall
\maketitle

\section{Introduction}\label{sec:Introduction}
\Gls*{ML} algorithms are increasingly being used across scientific research. In particular, these methods are being actively explored as research tools in condensed matter physics~\cite{Carleo2019machine, Carrasquilla2020machine, Bedolla2021machine, Johnston2022perspective}. One alluring application is the acceleration of many-body simulations, where one trains low-cost surrogate models to replace computationally intensive components of the simulation. Two notable examples are \gls*{ML} interatomic potentials for force field predictions in molecular dynamics simulations~\cite{Jinnouchi2019onthefly, Noe2020machine, Prasnikar2024machine} and \gls*{SLMC} algorithms~\cite{Liu2017selflearning, pan2025selflearning}, where \gls*{ML} surrogates are used to effectively sample configuration space~\cite{Huang2017accelerated, Liu2017selflearning, Nagai2017selflearning, Shen2018selflearning, Chen2018symmetry, Li2019accelerating, bakshi2026machine, pan2025selflearning}.  

A particularly important class of many-body algorithms is quantum embedding methods, 
like \gls*{DMFT} or its cluster extensions~\cite{Georges1996dynamical, Maier2005quantum}. These methods have been instrumental in shaping our early understanding of strongly correlated phenomena like the Mott transition~\cite{Rozenberg1994Mott-Hubbard, Bulla2001Finite-Temperature, Jaewook2001quantum}, unconventional superconductivity, and pseudogap behavior~\cite{Maier2006structure, Gull2013superconductivity, Dong2022quantifying, Mai2022intertwined, Abhishek2025cluster}. They are being used increasingly in \textit{ab initio} electronic structure calculations of strongly correlated materials~\cite{Kotliar2006Electronic} and systems driven out of equilibrium~\cite {Aoki2014nonequilibrium}. Given the importance and computational expense of these methods, there has been significant interest in \gls*{ML} acceleration~\cite{Nagai2017selflearning, Arsenault2014machine, rogers_bypassing_2021, Sheridan2021datadriven, Egor2024predicting, Lee2025language}.

\begin{figure*}[t]
     \centering
         \includegraphics[width=0.90\linewidth]{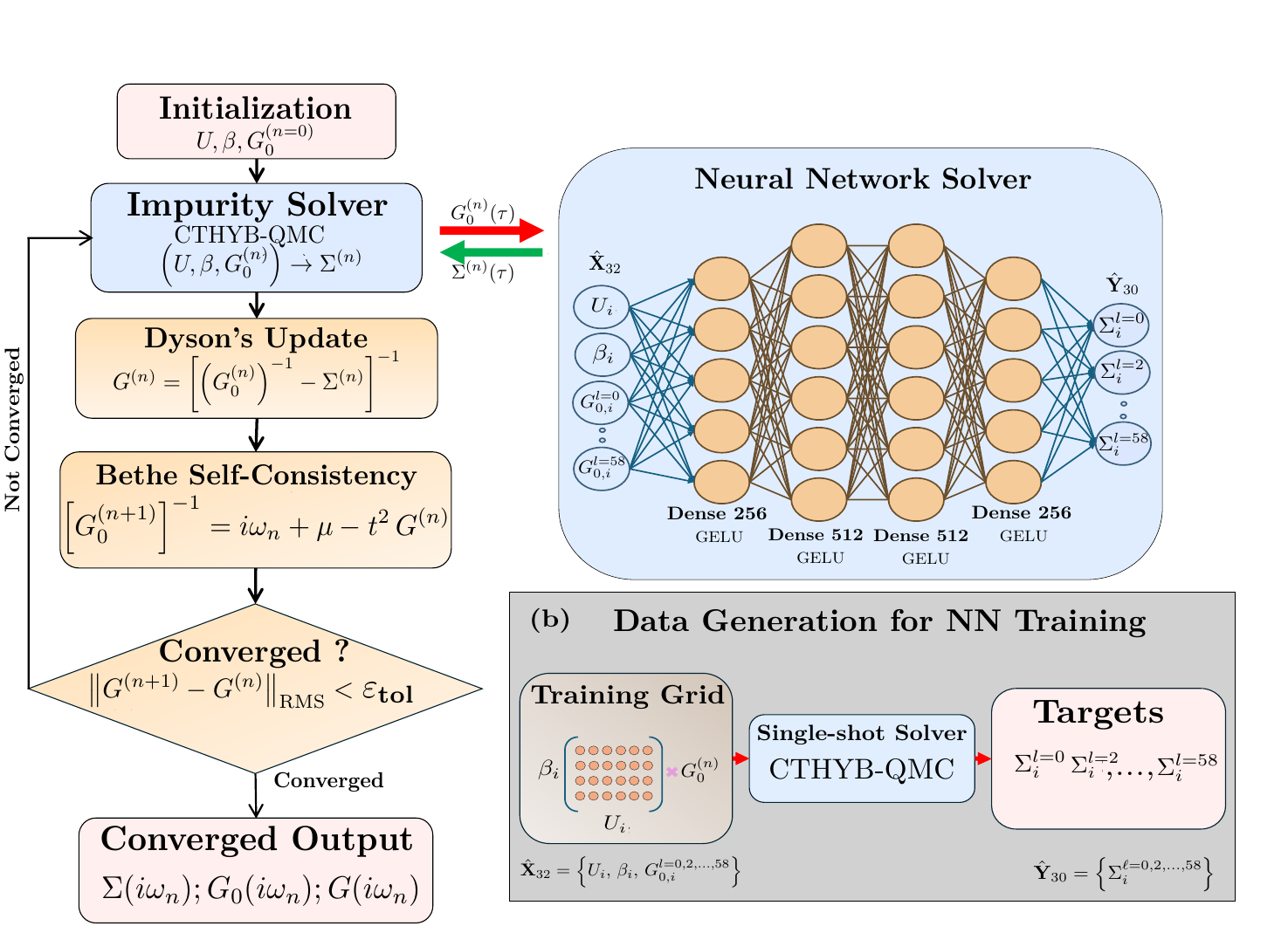}
     \caption{ (a) Left: The \gls*{DMFT} self-consistency loop. The computationally expensive part is the many-body calculation used to solve the impurity problem. Our goal is to replace this solver with an efficient \gls*{ML} architecture. Right: The \gls*{NN} used to replace the impurity solver in the \gls*{DMFT} loop consists of $4$ fully-connected dense layers of neurons with GELU activations. It accepts the Weiss mean-field Green's function $G_0(\tau)$, with the inverse temperature $\beta$ and the Hubbard interaction strength $U$ as input features, and predicts the impurity site's self-energy $\Sigma(\tau)$. (b) Illustration of the data generation by the single shot impurity \gls*{CTQMC} solver for a fixed $(U_{i},\beta_{i})$ grid for $10$ synthetic Weiss fields $G_{0}^{n}$ generated by finite pole expansion and the corresponding self-energies $\Sigma^{(n)}(\tau)$, both in terms of their Legendre coefficients $G_{0, i}^{l=0,2,\dots,58}$ and $\Sigma_{i}^{l=0,2,\dots,58}$, respectively, forming the features and target pairs for training of the \gls*{NN} $(\hat{{X}}_{32}, \hat{{Y}}_{30})$.}
     \label{fig:DMFT_loop}
\end{figure*}

The \gls*{DMFT} algorithm is shown in Fig.~\ref{fig:DMFT_loop}(a). It starts with an initial guess for the bath parameters encoded in the effective-medium Green's function $G_0(\tau)$. An accurate method like \gls*{QMC} is then used to compute the interacting Green’s function $G(\tau)$ of a quantum impurity problem defined by $G_0(\tau)$. The resulting impurity Green's function $G(\tau)$ is then used to update $G_0(\tau)$ in the next iteration, and this process is iterated until a self-consistency condition is reached.  
The computationally expensive part is the ``impurity solver,'' which calculates $G(\tau)$ for the interacting impurity problem. Traditionally, numerically exact methods like \gls*{CTQMC}~\cite{Gull2011Continuous}, \gls*{ED}~\cite{Liebsch2012temperature}, or \gls*{NRG}~\cite{Bulla2008numerical} are used; however, these can be computationally demanding, particularly at low temperatures and when generalized to multi-orbital problems. For example, popular \gls*{CTQMC} solvers scale as the cube of the inverse temperature and number of orbitals $O(N^3\beta^3)$~\cite{Gull2011Continuous}, making them costly at low temperatures and on large systems. Conversely, \gls*{ED}-based solvers discretize the mean-field bath and face an exponential growth of the Hilbert space when solving the many-body impurity problem. 

As mentioned, various \gls*{ML} approaches have been explored to accelerate the \gls*{DMFT} loop. For example, Ref.~\onlinecite{Arsenault2014machine} trained a \gls*{NN} model to predict 
correlation functions and quasiparticle weights for the three-dimensional Hubbard model using the initial hybridization function as an input. Similarly, Sheridan \textit{et al}.~\cite{Sheridan2021datadriven} built a \gls*{ML} solver for the \gls*{AIM} framed as a $\Delta$-learning problem, where the \gls*{NN} learned to predict the error corrections to computationally inexpensive estimates for the impurity Green's function. More recently, Lee \textit{et al}.~\cite{Lee2025language} expanded on this approach using a modified transformer architecture to map low‑cost Matsubara Green’s functions to high‑accuracy impurity Green's functions. Agapov \textit{et al}.~\cite{Egor2024predicting} trained a \gls*{NN} on $2{,}000$ training examples to predict $G(\tau)$ for two-dimensional lattices. They argued that their approach could replace the impurity solver within the \gls*{DMFT} loop but found it faced some difficulties in converging to accurate solutions without fine-tuning.

In general, these prior studies have found that appropriately trained \gls*{ML} surrogates could indeed replace more expensive numerically exact impurity solvers and reduce the time to solution. However, in all cases they used relatively large training datasets that are expensive to generate. This aspect poses challenges for generalizing this approach to multi-orbital models and, in particular, cluster extensions of \gls*{DMFT}, where generating sufficient amounts of high-quality training data will be computationally demanding. 

Motivated by this, here we examine the extent to which a \gls*{NN} can be \textit{efficiently} trained as a low-cost but accurate surrogate for the impurity problem in a \gls*{DMFT} calculation. Focusing on the half-filled Hubbard model, we show that a well-trained \gls*{NN}, trained on a fairly feasible 500 samples synthetic dataset, can predict the impurity self-energy $\Sigma^{(n)}(\tau)$ (for a given iteration $n$), from which the impurity Green's function $G^{(n)}(\tau)$ is obtained via Dyson's equation, using Hubbard $U$, inverse temperature $\beta = 1/T$ ($k_\mathrm{B} = 1$), and $G_0^{(n)}(\tau)$ as input features. 

Within the parameter regime well covered by the training data, the ML surrogates are sufficiently accurate to fully replace traditional \gls*{CTQMC} impurity solvers, thereby yielding dramatic acceleration. Crucially, this level of performance can be achieved with a relatively small training dataset. We also show that the model extrapolates well for a range of temperatures below the training set coverage. As expected, model errors can become significant with extrapolation; nonetheless, we find that imperfect model predictions can be useful as an initial guess that accelerates the convergence to self-consistency within traditional \gls*{DMFT} solvers.

\section{Methods}\label{sec:Methods}
Throughout this work, we consider the half-filled Hubbard model 
\begin{equation}\label{eq:H}
    \hat{H} = -t\sum_{\langle i,j\rangle,\sigma} \hat{c}^\dagger_{i,\sigma} \hat{c}^{\phantom\dagger}_{j,\sigma} + U \sum_i \left(\hat{n}_{i,\uparrow}-\tfrac{1}{2}\right)\left(\hat{n}_{i,\downarrow}-\tfrac{1}{2}\right), 
\end{equation}
defined on the Bethe lattice within \gls*{DMFT} (Fig. \ref{fig:DMFT_loop}). 
Here, $\hat{c}^\dagger_{i,\sigma}$ ($\hat{c}^{\phantom\dagger}_{i,\sigma}$) creates (annihilates) a spin-$\sigma$ electron on site $i$, $\hat{n}_{i,\sigma} = 
\hat{c}^\dagger_{i,\sigma}\hat{c}^{\phantom\dagger}_{i,\sigma}$ is the number operator, $t$ is the nearest-neighbor hopping, and $U$ is the Hubbard repulsion. 

In the first iteration, the Weiss mean-field $G_0^{(n)}(i\omega_n)$ is given by an integral
\begin{equation}
    G^{(n=0)}_0(i\omega_n) = \int_{-\infty}^{\infty} \frac{\rho(\varepsilon)}{i\omega_n +\mu-\varepsilon} d\varepsilon\,,
\end{equation}
where $\rho(\varepsilon)=\frac{1}{2\pi t^{2}}\sqrt{4t^{2}-\varepsilon^{2}}$ is the semi-circular density of states of the Bethe lattice. 
For a given $G_0^{(n)}(i\omega_n)$, each \gls*{DMFT} iteration calculates the self-energy $\Sigma^{(n)}(i\omega_n)$, which is a functional of $G_0^{(n)}(i\omega_n)$, $\beta$, and $U$, by solving the \gls*{AIM}. 
The local Green's function $G^{(n)}(i\omega_n)$ is then updated using the Dyson equation
\begin{equation}\label{eq:Dyson}
    G^{(n)}(i\omega_n)
    =
    \left[\left[G_0^{(n)}(i\omega_n)\right]^{-1}
    -
    \Sigma^{(n)}(i\omega_n)
    \right]^{-1}. 
\end{equation}
The updated Weiss field for the next iteration is determined by the self-consistency condition, which, for the Bethe-lattice, simplifies to \cite{Georges1996dynamical}
\begin{equation}\label{eq:Bethe_reduced}
    \left[G_0^{(n+1)}(i\omega_n)\right]^{-1}
    =
    i\omega_n + \mu - t^2\,G^{(n)}(i\omega_n). 
\end{equation}
This process is iterated until the \gls*{RMSE} 
\begin{equation}\label{eq:conv}
    \varepsilon_G^{(n+1)}
    =
    \left\| G^{(n+1)}(\tau)-G^{(n)}(\tau)\right\|_{\rm RMS}
    <
    \varepsilon_{\rm tol}.
\end{equation}
falls below a solver-dependent tolerance $\varepsilon_{\rm tol}$.

We used the \gls*{TRIQS} library~\cite{TRIQS} with the \gls*{CTHYB} \gls*{QMC} solver to solve the \gls*{AIM} in the \gls*{DMFT} loop. For each simulation, we use $ n_{\rm warmup}= 5\times10^{4}$ thermalization sweeps followed by $ n_{\rm cycles}=2\times 10^{5}$ measurement sweeps 
with a cycle length of $\ell_{\rm cyc}=1{,}200$. The cycle length was chosen to maintain a measured autocorrelation time between 0.5 and 1.5, as described in the \gls*{TRIQS} documentation~\cite{TRIQS}.

The Matsubara Green's functions $G_0(i\omega_n)$ and $G(i\omega_n)$ and self-energy $\Sigma(i\omega_n)$ are obtained through a Fourier transform from the corresponding functions on the imaginary time ($\tau$) axis.  
We adopt a compact Legendre polynomial basis \cite{Boehnke2011orthogonal} to represent all imaginary-time quantities 
\begin{equation}\label{eq:Legendre_P}
    F(\tau) = \sum_{l\,\mathrm{even}}^{l_{\rm max}} f_{l} \, P_{l}\!\left(\frac{2\tau}{\beta} - 1\right),
\end{equation}
where $F(\tau)$ can be the Green's function $G(\tau)$ or the self-energy $\Sigma(\tau)$, $P_{l}(\tau)$ are the $l$\textsuperscript{th} order Legendre polynomials defined on the interval $[-1,1]$, $f_{l}$ are the expansion coefficients, and $l_\mathrm{max}=60$ is a cut-off in the expansion order. Working at half-filling, particle-hole symmetry demands that $F(\tau) = F(\beta-\tau)$, which results in vanishing coefficients $f_l$ for odd $l$ in the expansion. Thus, both $G_{0}(\tau)$ and $\Sigma(\tau)$ are fully encoded by their 30 even-indexed coefficients  $\{f_0, f_2, \ldots, f_{58}\}$. In practice, we have found that taking $l_\mathrm{max} = 60$ is sufficient to obtain reliable results for all values of $U$ and $\beta$ considered here.

The architecture of our \gls*{NN} is shown in Fig.~\ref{fig:DMFT_loop}. It accepts the Hubbard interaction strength $U$, the inverse temperature $\beta$, and the 30 even Legendre coefficients of the
Weiss field ${G}_0(\tau)$ as a 32-dimensional input vector, and returns the 30 even Legendre coefficients of the self-energy $\Sigma(\tau)$ as an output vector. The network is comprised of four fully-connected layers of neurons with GELU activations and trained to minimize the physics-informed loss function
\begin{equation}\label{eq:loss}
    \mathcal{L}
    = \left\langle
        \sum_{l\,\mathrm{even}} w_l
        \bigl(f_l - \hat{f}_l\bigr)^2
      + \lambda_{\rm dec}
        \sum_{l\,\mathrm{even}} l^2\,\hat{f}_l^2
      \right\rangle_{\mathrm{batch}}.
\end{equation}
Here $\hat{f}_l$ denotes the predicted coefficient, $\lambda_{\rm dec} =10^{-4}$ controls the Legendre-decay regularization, and $\langle\cdot\rangle_{\mathrm{batch}}$ denotes the batch average. The per-coefficient weights $w_l$ are chosen to strongly penalize errors in the lower order coefficients, which carry the dominant spectral weight and were determined empirically through iterative refinement by comparing single-step \gls*{NN} predictions against \gls*{CTQMC} self-energies $\Sigma(\tau)$ across a range of weight configurations. The Legendre-decay term acts as a mild regularizer, penalizing unphysically large high-$l$ coefficients, enforcing the expected physical suppression of the higher order Legendre models in smooth imaginary-time functions. 

To impose the analytically exact constraint that the self-energy vanishes for $U = 0$, the training set is augmented with 100 synthetic samples at $U = 0$ with all target coefficients set to zero. 
(These training examples can be generated with no computational cost.) The network is optimized with AdamW~\cite{loshchilov2019decoupled} with a cosine learning rate schedule decaying from $\eta_0 = 10^{-3}$ over $10^4$ steps and weight decay $5\times10^{-4}$, for up to 2000 epochs with a batch size of $12$ and with early stopping whenever the validation loss fails to improve for $25$ consecutive epochs.

To generate the training data, we use synthetic data for the Weiss mean-field $G_0(\tau)$ and, together with $U$ and $\beta$ as inputs, compute the corresponding self-energy $\Sigma(\tau)$ using the \gls*{CTQMC} impurity solver. To ensure generalization across arbitrary bath geometries, we construct the Weiss field $G_{0}(i\omega_n) = [i\omega_n+\mu-\Delta(i\omega_n)]^{-1}$ by finite pole expansion of the corresponding hybridization function
\begin{equation}\label{eq:bath}
    \Delta(i\omega_n)
    = \sum_{p=1}^{N_p}\sum_{\sigma=\pm 1} \frac{|V_p|^2}{i\omega_n +\sigma \epsilon_p},
\end{equation}
with the number of poles $N_p\in [1,10]$, pole positions $\epsilon_p/t \in [0.05, 4.0]$, and hybridization strengths $V_p/t\in [0.1,0.5]$ drawn randomly from uniform distributions. Eq.~\eqref{eq:bath} explicitly respects particle-hole symmetry; generalization to doped models will require removing the symmetric sum over $\pm\epsilon_p$. The resulting $\Sigma(\tau)$ from \gls*{CTQMC} solver defined by its Legendre coefficients $\Sigma^{l=0,2,\dots 58}$ constitutes the training target.

The training samples were generated on a grid of $(U,\beta)$ values, with $10$ Weiss fields drawn at each grid point via the bath parametrization Eq.~\eqref{eq:bath}. 
[The $(U,\beta)$ grid is indicated by the white squares in Fig.~\ref{fig:training_data}.] 
Samples for which the Legendre-reconstructed $\Sigma(\tau)$ exhibited unphysical oscillations or sign changes were discarded, as these were taken to indicate insufficient Monte Carlo statistics, yielding a dataset of $N_\mathrm{train}=500$ training examples. In future studies, training data may be more expensive to generate and collect; accordingly, we intentionally limit the model here to 500 samples as a test of data efficiency. In Sec.~\ref{sec:dataset_size} we will also examine improvements in our model predictions for $N_\mathrm{train}=1600$ samples. 

We have found that the \gls*{NN} predicts impurity Green's functions very accurately for parameters within the coverage of the training data, effectively interpolating between training examples. However, we also examine the network's performance in regions of parameter space where it must extrapolate to lower temperatures than those represented in the training set. In the following sections, we refer to these two regions of parameter space as the ``interpolation'' and ``extrapolation'' regions of the phase diagram.  

\section{Results}\label{sec:Results}
\subsection{Performance of the solver}
\begin{figure}[t]
    \centering
    \includegraphics[width=\linewidth]{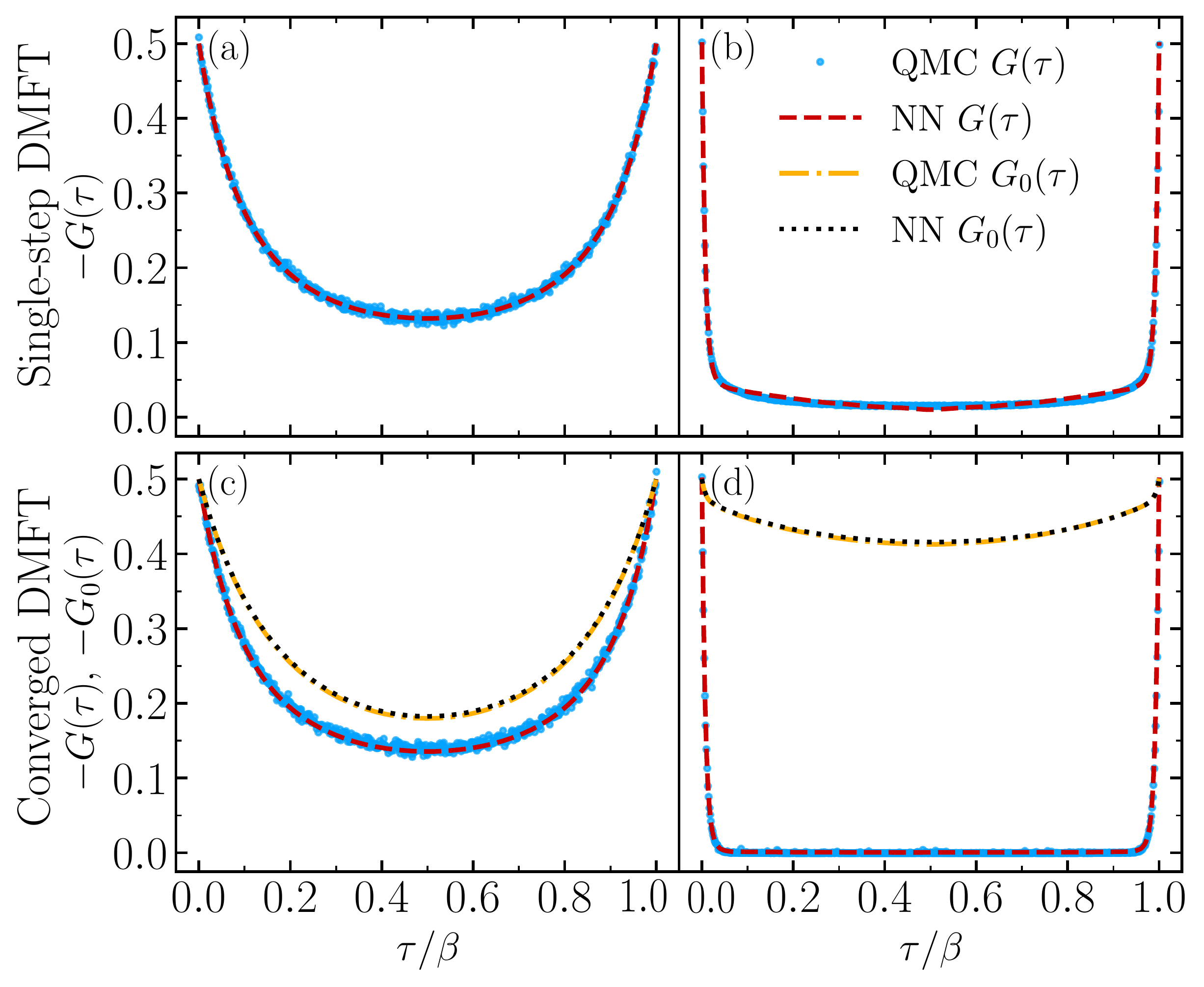}
    \caption{Imaginary-time Green's functions $G(\tau)$ for (a),(b) a single \gls*{DMFT} iteration and (c),(d) fully converged solutions. Results are shown for parameters in the metallic ($U/t = 2.50$, $\beta t = 6$, left column)  and insulating ($U/t = 9.0$, $\beta t= 30$, right column) regions of the phase diagram. We obtain excellent agreement between the \gls*{CTHYB} QMC solver and the \gls*{NN} solver; the \gls*{NN} solver reproduces both $G(\tau)$ and $G_{0}(\tau)$ for the fully converged solutions,  with an RMSE of $\mathrm{10}^{-3}$ at convergence. Both solvers achieve convergence in a comparable number of \gls*{DMFT} iterations; the QMC solver, however, required a runtime of \(\sim 2.1\times10^{3}~\mathrm{s}\) compared to the \gls*{NN} solver, which converged in $\sim 1.6\times10^{-1}~\mathrm{s}$, both timed on a single Intel Xeon Gold $6248R$ core ($3.00~\text{GHz}$) on the ISAAC cluster~\cite{ISAAC}.}
    \label{fig:Gtau}
\end{figure}

Figures~\ref{fig:Gtau}(a) and~\ref{fig:Gtau}(b) assess the accuracy of our \gls*{NN} in performing a single iteration of the \gls*{DMFT} loop. Figure~\ref{fig:Gtau}(a) shows the results for $U/t = 2.5$ and a relatively low inverse temperature $\beta t= 6$, while Fig.~\ref{fig:Gtau}(b) shows results for $U/t = 9.0$ and $\beta t= 30$. In both cases, the \gls*{NN} takes $G_0(\tau)$ as input (green dotted line) and predicts $\Sigma(\tau)$, from which $G(\tau)$ (red dashed line) is reconstructed. The \gls*{NN}-based predictions of $G(\tau)$ are in excellent agreement with those obtained using the more costly \gls*{CTHYB}-\gls*{QMC} solver (blue points), with the prediction generally falling within the \gls*{QMC} statistical error. We have obtained similarly consistent results for other parameter values within the interpolation region of the training set, demonstrating that our relatively small synthetic training dataset is sufficient for the learning task.  

\begin{figure}[t]
    \centering
    \includegraphics[width=\linewidth]{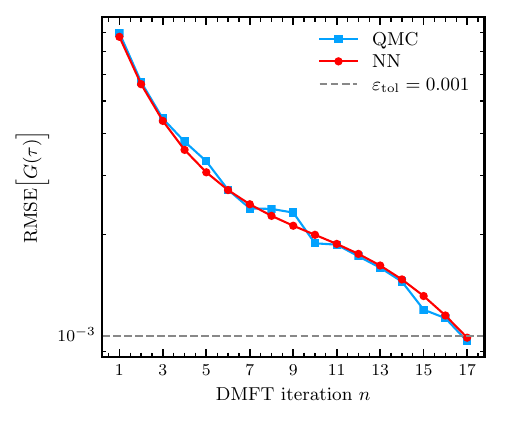}
    \caption{A comparison of the convergence of the \gls*{DMFT} loop using  
    either the \gls*{QMC} (\gls*{CTHYB}) solver or its low-cost surrogate \gls*{NN}. Both solvers were run for $U/t = 5$ and $\beta t = 20$ and converged to root-mean-square error (RMSE) $\epsilon \le \epsilon_\mathrm{tol} = 0.001$ in 17 iterations. 
    }
    \label{fig:Convergence}
\end{figure}

Figures~\ref{fig:Gtau}(c) and ~\ref{fig:Gtau}(d) compare the results of running the \gls*{DMFT} self-consistency loop to convergence using the low-cost \gls*{NN} surrogate with those obtained using the \gls*{QMC} solver for the same task. The data in panels (c) and (d) are for the same parameters used in panels (a) and (b), respectively. In both cases, we show $G_0^{(n-1)}(\tau)$ and the impurity Green's $G^{(n)}(\tau)$ for the final step of the \gls*{DMFT} loop after converging to a \gls*{RMSE} $\epsilon < 0.001$. The surrogate \gls*{NN} impurity solver produces a solution consistent with the \gls*{QMC} solver in both the metallic [Fig.~\ref{fig:Gtau}(c)] and insulating [Fig.~\ref{fig:Gtau}(d)] cases, demonstrating that the \gls*{NN} can obtain reliable solutions to the impurity problem. Importantly, Fig.~\ref{fig:Convergence} shows that the convergence rates for the two solvers are also similar, demonstrating that the \gls*{NN} reproduces the iterative fixed-point behavior of the \gls*{QMC}-based impurity solver. However, converging the solution using the \gls*{NN} solver (once trained) is approximately four orders of magnitude faster than the \gls*{QMC} solver.  

\subsection{Observables}

\begin{figure}[t]
    \centering
    \includegraphics[width=\linewidth]{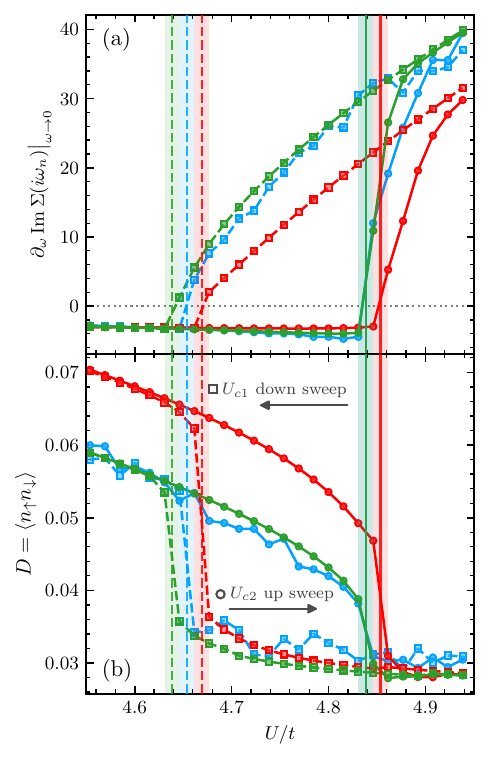}
    \caption{The evolution of (a) the low-frequency slope in the imaginary part of the self-energy $\Sigma(i\omega_n)$ and (b) double occupancy on the impurity site as a function 
    of $U/t$ at a fixed $\beta t = 25$. Results are shown for the converged solutions 
    obtained using a \gls*{QMC} (blue) and \gls*{NN} (red $N_\mathrm{train}=500$)  and (green $N_\mathrm{train}=1600$) impurity solvers. Two curves are shown in each case. One is obtained starting with an insulating solution and sweeping downward in $U/t$ to estimate $\UcI$ ($\square$) while the other starts with a metallic solution and sweeps upward in the interaction to obtain $\UcII$ ($\bigcirc$).}
    \label{fig:hysteresis}
\end{figure}

\begin{figure}[t]
    \centering
    \includegraphics[width=1.08\linewidth]{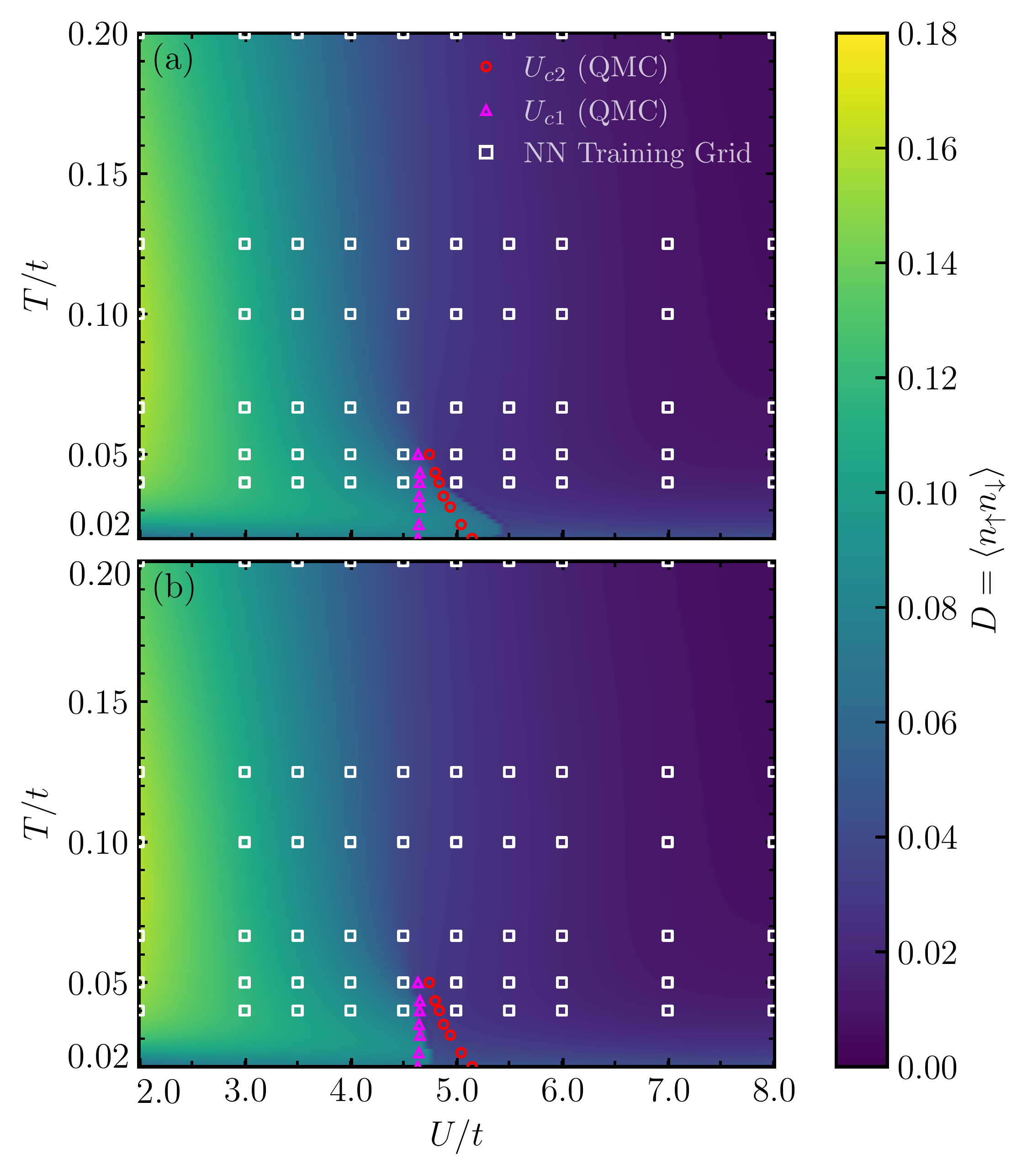}
    \caption{Double occupancy $D = \langle n_\uparrow n_\downarrow \rangle$ phase diagram tracked by the \gls*{NN} across uniform $10^{4}$ phase points in the $(U/t,\, T/t)$ grid by sweeping upwards in $U/t$ for (a) metallic and downwards for the (b) insulating branch. Circles and triangles denote the \gls*{QMC} spinodal points $\UcII$ and $\UcI$, respectively, extracted from the sign change in the low-frequency self-energy slope (see Fig.~\ref{fig:hysteresis}). The hollow white squares represent the \gls*{NN} training grid. The qualitatively accurate results of the \gls*{NN} for the phase diagram and coexistence region signify the interpolation and extrapolation capabilities of the \gls*{NN} impurity solver.} 
    \label{fig:training_data}
\end{figure}

The paramagnetic Mott metal-insulator transition in the half-filled Hubbard model is a well-characterized phenomenon within \gls*{DMFT}, having been extensively studied using different impurity solvers~\cite{Georges1996dynamical, Rozenberg1994Mott-Hubbard, Kotliar2002Compressibility, Bulla2001Finite-Temperature, Jaewook2001quantum, Bluemer2003mott}. \gls*{DMFT} predicts the transition to be first order at finite temperatures, with a coexistence region bounded by the spinodals $\UcI$ and $\UcII$, below (above) which the insulating (metallic) solution ceases to exist. To examine the reproducibility of this coexistence region by the \gls*{NN}, we determine the upper metallic $\UcII$ and lower insulating $\UcI$ boundaries, which can be estimated using the slope in the self-energy $\Sigma(i\omega_n)$ at the lowest Matsubara frequency. 

Figure~\ref{fig:hysteresis}(a) plots the low-frequency Matsubara-axis self-energy slope,
\begin{equation}\label{eq:slope}
    \left.\frac{\partial\mathrm{Im}\,\Sigma(i\omega_{n})}{\partial\omega}\right|_{\omega\rightarrow 0} \approx \frac{\mathrm{Im}\,\Sigma(i\omega_1) 
    - \mathrm{Im}\,\Sigma(i\omega_0)}{\omega_1 - \omega_0},
\end{equation}
evaluated using the two lowest Matsubara frequencies, $\omega_{0,1}=\pi/\beta,\,3\pi/\beta$, as a function of $U/t$ for fully converged \gls*{DMFT} solutions. Figure~\ref{fig:hysteresis}(b) plots the corresponding impurity-site double occupancy, computed using the Galitskii-Migdal
relation~\cite{GalitskiiMigdal1958_ZhETF,GalitskiiMigdal1958_JETP,Gull2012,vanLoon2016} 
\begin{equation}\label{eq:Docc}
  D = \langle n_\uparrow n_\downarrow \rangle = \tfrac{1}{4}
    + \frac{1}{U\beta}\sum_{n}
      \mathrm{Re}\bigl[\Sigma(i\omega_n)\,G(i\omega_n)\bigr],
\end{equation}
where the factor of $1/4$ arises from our Hartree-subtracted convention for the self-energy at half-filling. 
In both cases, we compare results with the \gls*{NN} (red) and \gls*{CTQMC} (blue) impurity solvers. Circle and square symbols denote calculations initialized from metallic and insulating solutions, respectively, with interaction sweeps performed upward and downward in $U$. We observe that both observables locate the Mott metal-insulator transitions consistently for the two solvers. The spinodals $\UcI$ and $\UcII$ are determined by the sign change in the low-frequency Matsubara-axis slope in Fig.~\ref{fig:hysteresis}(a) on insulating and metallic branches, respectively. These values are corroborated by the double occupancy hysteresis shown in Fig.~\ref{fig:hysteresis}(b), yielding a consistent characterization of the coexistence region.

The phase boundaries obtained using the \gls*{NN} solver are in close agreement with those extracted using the \gls*{CTQMC} solver. Nevertheless, a small but systematic offset is observed in the double occupancy
along the metallic branch, as shown in Fig.~\ref{fig:hysteresis}(b), where $\langle n_{\downarrow}n_{\uparrow}\rangle$ is slightly overpredicted by the \gls*{NN}. This discrepancy likely stems from small inaccuracies in the low-frequency self-energy entering Eq.~\eqref{eq:Docc} and is largely corrected by increasing the training set size to 1600 (see Sec.~\ref{sec:dataset_size}). By contrast, the extraction of $\UcI$ and $\UcII$ remains unaffected, since the low-frequency slope in $\mathrm{Im}\,\Sigma(i\omega_{n})$ is governed primarily by the relative difference between the first two Matsubara frequencies and is therefore considerably less sensitive to such shifts. 
 
\begin{figure}[t]
    \centering
    \includegraphics[width=\linewidth]{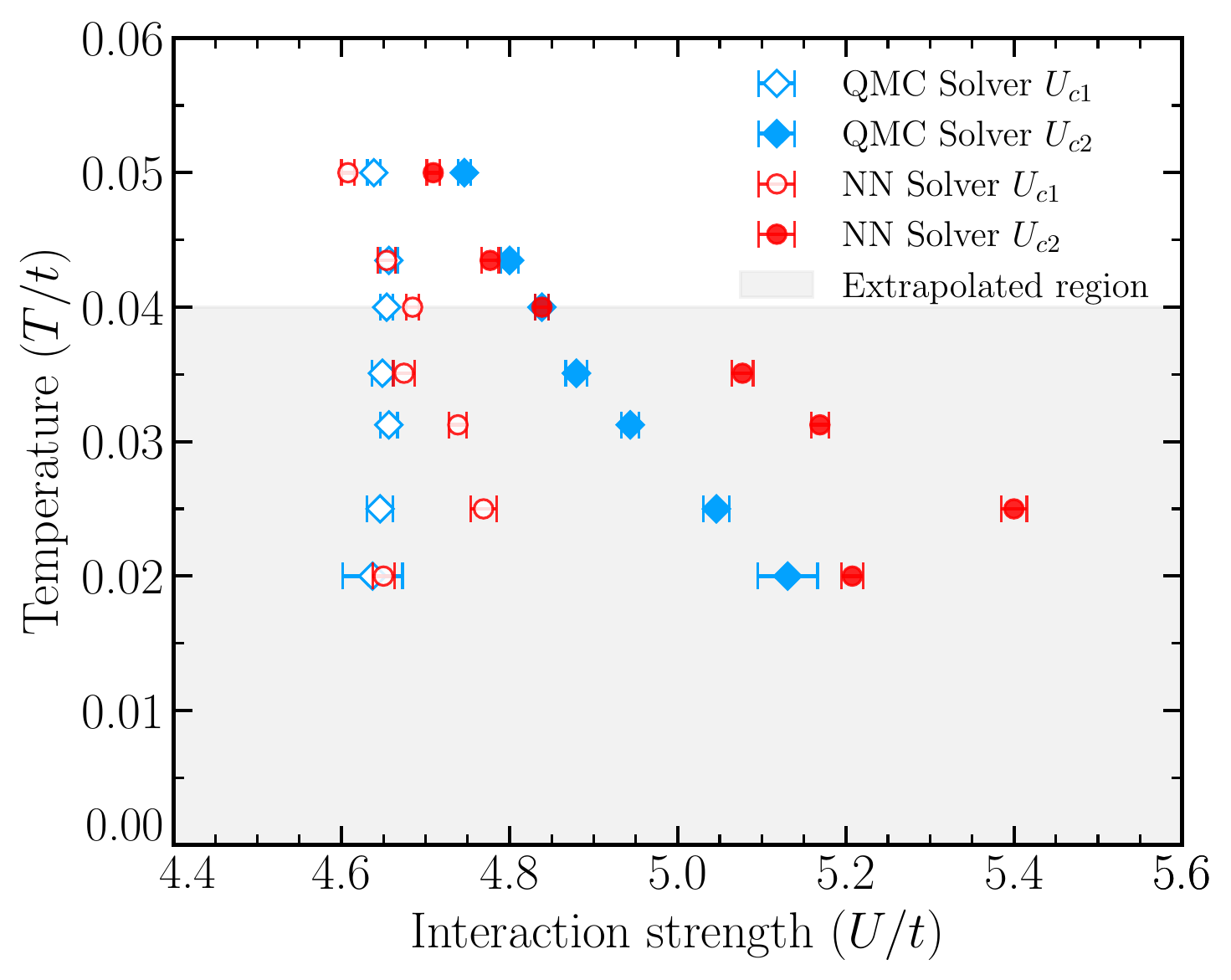}
    \caption{The phase diagram predicted for the singleband Hubbard model on the Bethe lattice using the \gls*{NN} impurity solver, focusing on the coexistence region in the $U$-$T$ plane. 
    The $\UcI$ and $\UcII$ estimates obtained from the change in slope of the imaginary part of the self-energy at the 0th Matsubara frequency [see Fig.~\ref{fig:hysteresis}(a)] are plotted and compared against estimates obtained using the \gls*{QMC} impurity solver. The white (gray) region 
    denotes the interpolation (extrapolation) regions of the phase space.}
    \label{fig:phase_diagram}
\end{figure}

We carried out a similar analysis across a range of parameters spanning both the interpolation and extrapolation regimes of the trained \gls*{NN}, and obtained the phase boundaries shown in Fig.~\ref{fig:phase_diagram}. The  \gls*{NN} reproduces the phase boundaries determined by the \gls*{CTQMC} solver well in the interpolation regime, and qualitatively in the extrapolation regime just outside of the training data coverage. Note, however, that the agreement worsens as one extends the extrapolations far beyond the interpolated regime.  

\subsection{Phase diagram and extrapolating beyond the training data}
To assess the interpolation and extrapolation capabilities of the \gls*{NN} impurity solver, we generated a phase diagram by mapping the double occupancy $D(U/t, T/t)$ [Eq.~\eqref{eq:Docc}] over a uniform $100\times100$ grid spanning $U/t \in [2.0, 8.0]$ and $T/t \in [0.02, 0.20]$. We performed independent phase scans with increasing and decreasing $U/t$, analogous to the hysteresis analysis shown in Fig.~\ref{fig:hysteresis}. The \gls*{NN} completes this dense scan of $10^{4}$ points in $\sim 8.3\times10^{3}~\mathrm{s}$. In sharp contrast, just one \gls*{CTHYB} DMFT iteration at a single point costs $\sim 2.1\times10^{3}~\mathrm{s}$, so a full $100\times 100$ scan requiring multiple iterations each would be prohibitive. 

The \gls*{NN} accurately interpolates across the trained $(U/t,T/t)$ grid, reproducing the key features of the metallic and insulating regimes, and yielding qualitatively correct predictions for the Mott coexistence region bounded by $\UcI$ and $\UcII$. Crucially, the lowest temperatures in this scan, $T/t \lesssim 0.04\,(\beta t \gtrsim 25)$, lie in the extrapolation regime; nevertheless, the \gls*{NN} continues to produce a well-resolved coexistence region that remains qualitatively consistent with the \gls*{QMC} spinodals $\UcI$ and $\UcII$, demonstrating its ability to generalize under moderate extrapolation in $\beta t$.

\subsection{Dataset Size Scaling}\label{sec:dataset_size}

\begin{figure}[t]
    \centering
    \includegraphics[width=\linewidth]{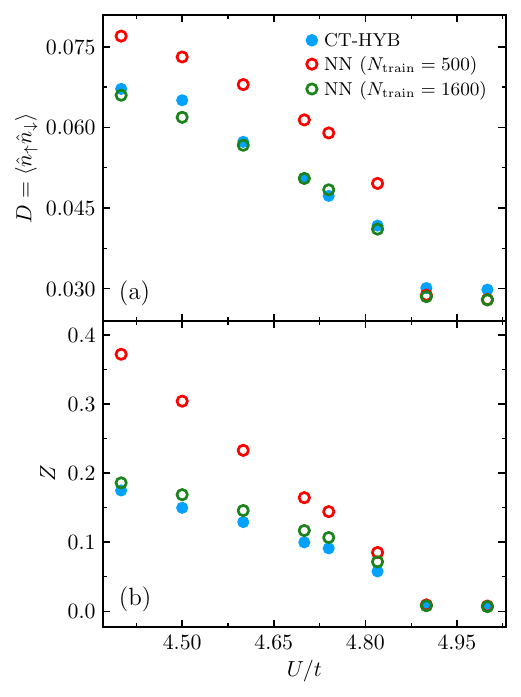}
    \caption{(a) Double occupancy $D=\langle n_{\uparrow}n_{\downarrow}\rangle$ and (b) quasiparticle weight $Z$ as a function of interaction strength $U/t$ along the metallic branch up-sweep at $\beta t=25$. Results are shown from fully converged \gls*{DMFT} for \gls*{CTHYB} solver (filled blue circles), \gls*{NN} surrogate solvers trained on $N_\mathrm{train}=500$ samples (open red circles) and $N_\mathrm{train}=1600$ (open green circles).} 
    \label{fig:scaling}
\end{figure}

\begin{figure}[t]
    \centering
    \includegraphics[width=\linewidth]{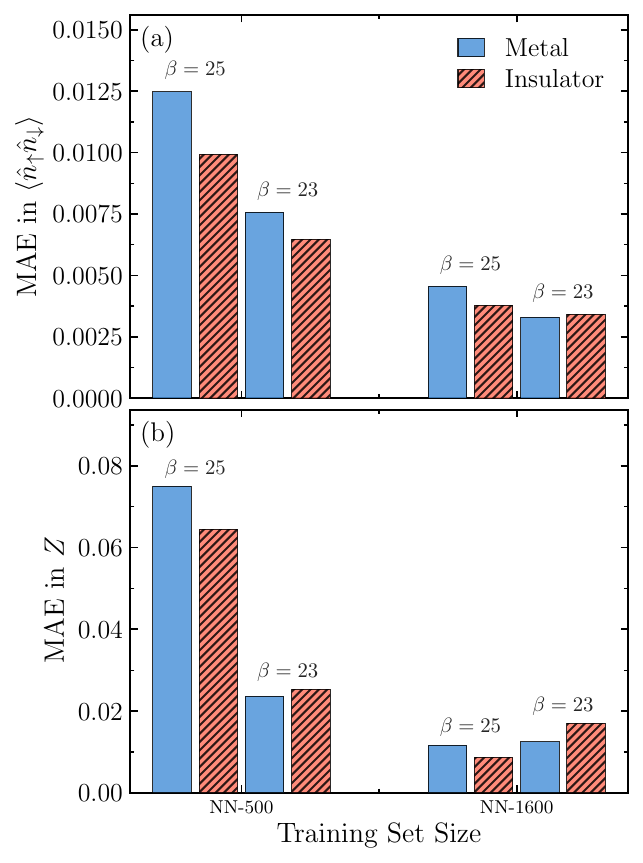}
    \caption{Mean absolute error (\gls*{MAE}) in the (a) double occupancy $D=\langle n_{\uparrow}n_{\downarrow}\rangle$ and (b) quasiparticle weight $Z$, resulting from converged \gls*{DMFT} solutions for the metallic and insulating sweeps averaged across $40$ $U/t$ points at ($\beta t = 25, 23$). Results for \gls*{NN} surrogates trained on (\gls*{NN}-$500$) and (\gls*{NN}-$1600$) samples are calculated with \gls*{CTHYB} solutions as reference for each branch.}
    \label{fig:mae_bars}
\end{figure}

Despite the qualitative accuracy of $N_\mathrm{train}=500$ model in predicting spinodals, we found a systematic overestimation of double occupancy $D=\langle n_{\uparrow} n_{\downarrow}\rangle$, particularly in the metallic branch. To assess the sensitivity of the model predictions to the size of the training dataset, we trained another \gls*{NN} surrogate with $N_\mathrm{train}=1600$ samples drawn from the same synthetic distribution. This newly trained model performs significantly better in predicting quantitatively correct observables, as shown in Fig.~\ref{fig:scaling}. Here, we show results for the double occupancy [Fig.~\ref{fig:scaling}(a)] and quasiparticle weight $Z \approx \left[1-\frac{\mathrm{Im}\Sigma (i\pi/\beta)}{\pi/\beta}\right]^{-1}$ [Fig.~\ref{fig:scaling}(b)] 
for the metallic sweep at $\beta t= 25$. 
To provide a more comprehensive analysis for the performance of the two \gls*{NN} impurity solvers, Fig.~\ref{fig:mae_bars} presents the mean absolute errors in the predictions in insulating (down) and metallic (up) sweeps across $40$ different $U/t$ values and for $\beta t=23$ and $25$. (Both temperatures lie in the interpolative regime.) The results show a consistent decrease in mean absolute errors for both observables as the training set size increases from $N_\mathrm{train}=500$ to $1600$. For example, the \gls*{MAE} in double occupancy at $\beta t= 25$ reduces by a factor of $2.7$ on the metallic sweep and $2.6$ on the insulating sweep. The improvement in predicting $Z$ is even better, with \gls*{MAE} dropping by $6.5$ in the metallic sweep and $7.5$ in the insulating sweep. The $N_\mathrm{train}=1600$ model achieves the \glspl*{MAE} below $0.005$ in $D=\langle n_{\uparrow} n_{\downarrow} \rangle$ and $0.012$ in $Z$, approaching \gls*{CTHYB} accuracy at a training cost that remains two orders of magnitude smaller than prior studies, which employed $16,000$ samples~\cite{valenti2026neural}.

\subsection{Accelerated solutions in the extrapolation region}
The results shown in Figs.~\ref{fig:training_data} and \ref{fig:phase_diagram} establish that a well-trained \gls*{NN} performs well when interpolating between points in its training dataset but less so when extrapolating to lower temperatures. In cases where performance decreases, it is interesting to ask whether its output can be efficiently refined using the exact solver. To test this, we ran the \gls*{DMFT} loop to convergence deep in the extrapolation regime using the fast \gls*{NN} solver. We then used the output of that calculation to seed a new \gls*{DMFT} loop, switching now to the more expensive \gls*{CTHYB} \gls*{QMC} solver. We refer to this hybrid approach as the \gls*{QMCacc} approach. 

\begin{figure}[t]
      \centering
  \includegraphics[width=\linewidth,height=0.65\textheight,keepaspectratio]{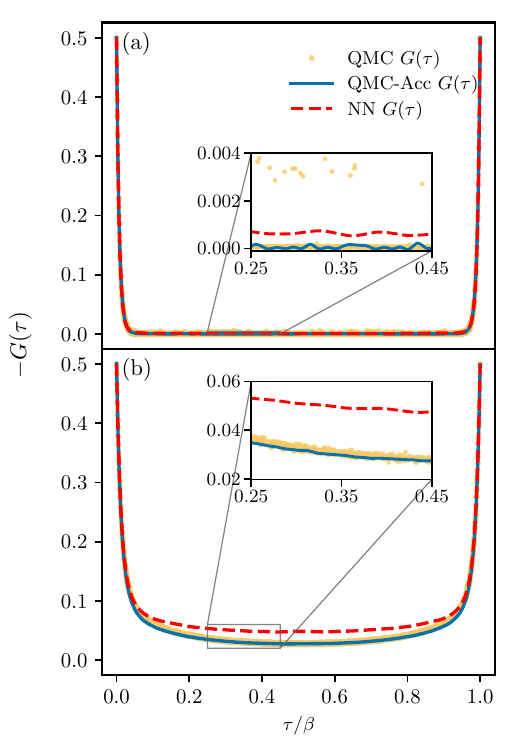}
  \caption{ Imaginary-time Green's functions $G(\tau)$ at $\beta t = 32$ (extrapolated region) for (a) insulating $(U/t=9.0)$ and (b) metallic $(U/t=4.5)$ regimes, comparing \gls*{CTHYB} \gls*{QMC}, \gls*{QMCacc}, and \gls*{NN} impurity solvers at converged \gls*{DMFT}. }
  \label{fig:QMCacc}
\end{figure}

Figure~\ref{fig:QMCacc} shows the results of this test for two parameter sets in the extrapolated regime ($\beta t = 32$) and on the insulating ($U/t = 9$, panel a) and metallic ($U/t = 4.5$, panel b) sides of the phase diagram. To quantify the agreement of the \gls*{QMCacc} and the \gls*{CTQMC} solutions we define a relative $L^{2}$ error in $\tau$-space $\varepsilon_{2,G(\tau)}$ given by  
\begin{equation}
    \varepsilon_{2,G(\tau)} = \left[ 
        \frac{\sum_{k} |G_{\text{QMC-Acc}}(\tau_{k})-G_{\text{QMC}}(\tau_{k})|^{2}}{\sum_{k} | G_{\mathrm{QMC}}(\tau_{k})|^{2}}
    \right]^{1/2}.
\end{equation}
In the insulating regime \gls*{QMCacc} reproduces the converged \gls*{CTQMC} solution with   $\varepsilon_{2,G(\tau)} = 1.77\times10^{-2}$ converging in $2$ \gls*{DMFT} iterations in 341 $\mathrm{s}$ compared with 6 iterations over 1,169 $\mathrm{s}$ for the pure \gls*{CTQMC}-based convergence, showing a $3.4\times$ reduction in the wall time. In the metallic regime, \gls*{QMCacc} achieves a $\varepsilon_{2,G(\tau)} = 4.35\times10^{-2}$ in $2$ iterations ($341~\mathrm{s}$) versus 5 iterations (1,930~$\mathrm{s}$) for \gls*{CTQMC} reducing the total computation time by $5.7\times$.

\section{Discussion \& Conclusions}\label{sec:Conclusions}
We have shown that a compact \gls*{NN}-based surrogate trained on a relatively small synthetic dataset can provide an accurate solution for a \gls*{DMFT} simulation of the half-filled single-band Hubbard model. 
In particular, our \gls*{NN} model reproduces the behavior of a \gls*{CTQMC} impurity-solver on an iteration-by-iteration basis, accurately capturing the converged \gls*{DMFT} solutions in the interpolation regime, and provides viable warm starts for exact solvers beyond that regime. 
The \gls*{NN}-surrogate predictions for the \gls*{DMFT} self-energy reproduce the impurity Green's function $G(\tau)$ in the metallic and insulating regions of the phase diagram, capturing the correct behavior for metallic and insulating initial seeds of the \gls*{DMFT} loop. 
The model also performs qualitatively well in the extrapolative regime for temperatures close to the training set boundary. This observation naturally led to the finding that the trained model can be used to initialize new simulations and accelerate their convergence. However, our relatively compact and efficient \gls*{ML} framework yields small discrepancies in some measured quantities, such as double occupancy, highlighting the sensitivity of the correlation functions stemming from our model prediction of the self-energy. 

Several aspects of the physical problem we study here and our workflow help to explain the accuracy of the \gls*{ML} model predictions. Starting from the half-filled Hubbard model on the Bethe lattice, we leveraged particle-hole symmetry to constrain the learned mapping and maintain the correct density, thereby removing potential complications arising from chemical potential tuning. Moreover, the Bethe lattice admits a particularly simple self-consistency condition, $\Delta(i\omega_{n})= t^{2}G_{\mathrm{loc}}(i\omega_{n})$, making the \gls*{DMFT} mapping smoother and plausibly easier for the \gls*{ML}-surrogate to learn. The Legendre polynomial basis parameterization is known~\cite{Boehnke2011orthogonal} to be an efficient representation of many-body Green's functions, and we find it to be similarly efficient for the self-energy. We note that our approach of using the Green's function $G_{0}(\tau)$ as an input, which efficiently encodes Hamiltonian parameters such as the lattice type, filling or chemical potential, bandwidth, and symmetries in a single function, allows for straightforward extensions to other impurity Hamiltonians and may have improved universality and transferability compared to other approaches which use the Hamiltonian parameters directly.  

It is important to emphasize that the \gls*{ML}-surrogate contains enough expressivity to capture essential features of the Mott transition, as exemplified by Fig.~\ref{fig:hysteresis}. The surrogate tracks the hysteresis curves arising from the choice of a metallic or insulating seed at the beginning of the simulation. Although some quantitative discrepancies in the double occupancy are clearly visible, we believe they can be mitigated with more training data and architectural design tweaks. 

The interpolative power of \gls*{ML}-based approaches is readily apparent in our predictions of $\UcI$ and $\UcII$ in Fig.~\ref{fig:phase_diagram}. High-fidelity results are achievable in this region, enabling high-density predictions in phase diagrams that are well-represented by the training data distribution. However, as is well known, high-dimensional \gls*{NN}-based solution manifolds are often unconstrained outside the training manifold, and extrapolations of these models to out-of-distribution regimes are typically unreliable. In our case, the \gls*{NN} model exhibits enough extrapolation power to capture schematic aspects of the unseen region, suggesting that the model could be fine-tuned by partially freezing weights and training on additional examples in this region. Moreover, the model can accelerate convergence by passing off the NN-converged solution to the exact impurity solver (as shown in Fig.~\ref{fig:QMCacc}), yielding trustworthy predictions in the extrapolative region of the phase diagram. In all cases, extra care is needed when additional competing phases and broken-symmetry channels may enter, either between or outside the grid of training examples. 

Compared to earlier \gls*{ML} studies of impurity solvers and \gls*{DMFT}~\cite{Arsenault2014machine, Sheridan2021datadriven, Egor2024predicting, Lee2025language}, our work deliberately targets compactness and data economy for the application at hand. In future studies of more complex models and solvers beyond the single-impurity problem, generating training data will become increasingly expensive. Accordingly, alongside questions of feasibility, representation, error-correction, and modern \gls*{ML} architectures, it is reasonable to expect that the number of training examples may remain small. Our results suggest that this approach is not only feasible but accurate enough to enable genuine exploration of challenging regimes in correlated-electron physics. While preparing this manuscript, Ref.~~\cite{valenti2026neural} appeared as a preprint. While this work is broader in scope and more materials-facing, our work uses roughly two orders of magnitude fewer training examples. This data efficiency may be related to our choice of learning target: rather than predicting the Green’s function $G(\tau)$ directly, our surrogate predicts the self-energy $\Sigma(\tau)$, from which $G(\tau)$ is recovered through Dyson’s equation. For the half-filled problem studied here, we found that learning the self-energy required fewer training examples than learning the Green’s function itself. One possible explanation is that learning $\Sigma(\tau)$ resembles a form of $\Delta$-learning, in which the known $G_{0}(\tau)$ provides useful scaffolding, thereby reducing error sensitivity and yielding a more favorable target for Legendre expansion.

Taken together, our results suggest a new paradigm for exploring many-body Hamiltonians with \gls*{ML}-based quantum impurity solvers. As mentioned above, carefully constructed models on small synthetic datasets can enable essentially continuous predictions of phase diagrams, and their utility can be extended to impurity solver accelerators. The newly acquired exact results may be reused for further self-improving training, thereby improving model accuracy and expanding its mapping to new physical regions of important model Hamiltonians. 


\section*{Acknowledgments}
This work was supported by the U.S.~Department of Energy, Office of Science, Office of Basic Energy Sciences, under Award Number DE-SC0022311. Notice of Copyright: This manuscript has been authored by UT-Battelle, LLC, under contract DE-AC05-00OR22725 with the US Department of Energy (DOE). The publisher acknowledges the US government license to provide public access under the DOE Public Access Plan (http://energy.gov/downloads/doe-public-access-plan).\\ 

\section*{Data Availability} The data supporting this study will be deposited in an online repository upon acceptance of the final version of the paper for publication. Until that time, the data will be made available upon reasonable request. 


\bibliography{references}

\end{document}